\documentclass[pra,aps,twocolumn,superscriptaddress]{revtex4}
\usepackage{amsfonts}
\usepackage{amsmath}
\usepackage{amssymb}
\usepackage{graphicx}
\usepackage{xcolor}
\usepackage{float}
\usepackage{ulem}
\usepackage{lipsum}
\usepackage{times}

\usepackage[colorlinks,breaklinks,linkcolor=blue,anchorcolor=blue,citecolor=blue,urlcolor=blue]{hyperref}

\begin{document}
\title{Narrowband nonreciprocal transmission in a Doppler-broadened multilevel atomic system}

\author{Ying-Xue Zhang}
\affiliation{College of Physics and Hebei Key Laboratory of Photophysics Research and Application,
                Hebei Normal University, Shijiazhuang, Hebei 050024, China}

\author{Hui-Min Zhao}
\email{hmzhao@hebtu.edu.cn}
\affiliation{College of Physics and Hebei Key Laboratory of Photophysics Research and Application,
                Hebei Normal University, Shijiazhuang, Hebei 050024, China}

\author{Yan-Kui Bai}
\email{ykbai@semi.ac.cn}
\affiliation{College of Physics and Hebei Key Laboratory of Photophysics Research and Application,
                Hebei Normal University, Shijiazhuang, Hebei 050024, China}

\date{\today}

\begin{abstract}
We theoretically investigate narrowband nonreciprocal transmission in a Doppler-broadened five-level M-type atomic medium.
After Doppler averaging, the M-type system retains a narrow near-resonant electromagnetically induced transparency window that is narrower than that of a standard three-level $\Lambda$ system.
This spectral narrowing is associated with the decreasing separation between the two inner dressed states as atomic velocity increases.
The transparency window exhibits pronounced direction-dependent transmission, with high forward transmission and strong suppression in the reverse direction.
The transparency linewidth and forward-backward probe transmission can be controlled via two independent driving fields, while the central transparency frequency is kept essentially unchanged.
The resulting narrowband nonreciprocity supports high forward transmission together with strong suppression in the reverse direction, which is suitable for frequency-selective optical filtering and nonreciprocal optical transmission in thermal atomic media.
Our scheme with these specific features is geared towards enabling compact vapor-cell-based nonreciprocal devices for quantum network applications.
\end{abstract}

\maketitle

\section{Introduction}
Optical nonreciprocity, characterized by the asymmetry propagation in forward and backward directions, is crucial for practical applications such as signal isolation~\cite{NatPhoton9.79,PRXQuantum2.030309}, optical routing~\cite{NP130.013601,PRL130.013601,PRA105.013711}, quantum network stabilization~\cite{NatPhys21.825}, invisible sensing~\cite{SciChinaPhysMechAstron67.240313}, low-noise information processing~\cite{NatElectron6.76}, and so on.
Various approaches have been proposed to realize optical nonreciprocity, including magneto-optical effects~\cite{JApplPhys124.103104,PRAppl24.034056,OL50.1552,OE32.17220},
atomic motion in thermal vapors~\cite{Optica13.1317,PRL125.123901,PRL121.203602,OptCommun462.125343,OptCommun489.126861,PRAppl12.054004,PRL123.033902},
moving optical lattices~\cite{PRA92.053859,OptComm410.916,PRL132.083202}, spatiotemporal modulation~\cite{PRL123.063901,NatCommun7.13657,NatPhoton11.774,PRA112.033522,PRL109.033901,PRA109.043103,PRL126.193901,PRL108.153901,NewJPhys27.033015}, and nonlinear optical interactions~\cite{PRA106.063523,PRA110.023723,PRL128.083604,OE.6284}.
More recent developments involve a variety of physical mechanisms and platforms, including topological photonics~\cite{PRL134.193602,Nanophotonics9.4473}, cavity-assisted architectures~\cite{OptCommun415.39,PRL123.233604,PRL134.196904,OptExpress29.25477}, optomechanical and acousto-optic interactions~\cite{NatCommun9.1798,NatCommun9.1797,OptExpress31.7120,Nature568.65,OptSpectrosc113,PRAppl10.064037,PRA103.053501}, chiral quantum optical systems~\cite{PhysRevA90.043802,PhysRevA103.063706,PhysRevA97.062318,Nature541.473,Science354.1577}, non-Hermitian schemes based on loss-induced asymmetry~\cite{LightSciAppl10.30,npjQuantumInf10.75,PhysRevLett134.176601}, and frequency-conversion processes~\cite{PhysRevA106.032606}.
Despite their diversity, most of these approaches operate in broadband regimes and provide only limited control over spectrally selective nonreciprocal responses.

In fields such as precision spectroscopy~\cite{Photonics12.203,PRL106.033001}, high-speed optical communication~\cite{PRL132.140802,PRAppl19.044079}, and quantum information processing~\cite{PhysRevLett.134.020802},
optical systems often need to be transmitted within a selected frequency band.
In these settings, nonreciprocal transmission can provide direction-dependent control over optical signals within the frequency band of interest.
Restricting this response to a narrow spectral window is therefore valuable when selectivity over frequency is required.
Narrowband optical responses can be achieved using spectral filtering or cavity enhancement~\cite{PRL134.053801,PRA110.033722}, but these approaches may require additional optical components and increase system complexity.
Atomic media offer an alternative route, as their intrinsic linewidths and resonant dispersion can support narrowband optical responses~\cite{PRA111.013103,PRA113.043702}.
In thermal vapors, however, Doppler broadening reduces this natural selectivity and weakens the coherence effects~\cite{PhysRevA51.576,PhysScr.97.025503}.
Multilevel coherence offers a means of controlling the absorption and dispersion responses, as illustrated by phenomena such as electromagnetically induced transparency (EIT)~\cite{PhysRevA.77.063807,PhysRevA.66.013805,Pramana2020}.
These features have also motivated the study of nonreciprocal transmission in atomic media.

The intrinsically narrow spectral features of atomic transitions are strongly broadened in thermal vapors by Doppler effects, which are naturally associated with broadband optical responses.
In the $\Lambda$-type configuration studied in Ref.~\cite{PRL125.123901}, strong collisional broadening produces a wide absorption profile, and the direction dependence arises from direction-dependent population trapping over the velocity distribution.
Consequently, the nonreciprocal response is broadband and robust against the precise one-photon detunings of the control fields.
Broadband nonreciprocity has also been demonstrated in a five-level $\Lambda$ system~\cite{NewJPhys.26.043018}:
by operating at large single-photon detunings, where the intermediate state is adiabatically eliminated and the response is governed by two-photon resonance, robust transmission asymmetry was achieved over hundreds of megahertz without sensitivity to fine one-photon spectral features.
This motivates a further question: can a narrow, direction-dependent transmission window be sustained after Doppler averaging, and how can its linewidth be tuned while maintaining high forward transmission and strong reverse suppression?
Wave-vector geometry provides one means of controlling Doppler sensitivity in multiphoton processes; for example, an appropriate three-photon geometry can cancel first-order Doppler shifts for all velocity classes in thermal vapors~\cite{PhysRevA.111.033112,CPL.43.010302}.
Unlike the broadband schemes discussed above, the present work focuses on nonreciprocity emerging near one-photon resonance.
By employing an M-type configuration, in which the highest excited state of earlier $\Lambda$ schemes~\cite{NewJPhys.26.043018} is replaced by a ground-state sublevel, multiple coherent driving fields generate a velocity-dependent dressed-state structure.
This structure gives rise to a narrow, direction-selective transparency window embedded within the Doppler-broadened absorption profile.
We analyze how assistant driving fields modify the velocity-dependent dressed-state structure and the corresponding susceptibility, allowing controlled access to the spectral pathways that influence transparency and directional response.
These capabilities support compact and thermally stable narrowband optical isolation and spectral filtering, relevant to quantum memory protection and directional quantum communication.

\section{Theoretical Model}
\begin{figure}[ptbh]
\includegraphics[width=0.4\textwidth]{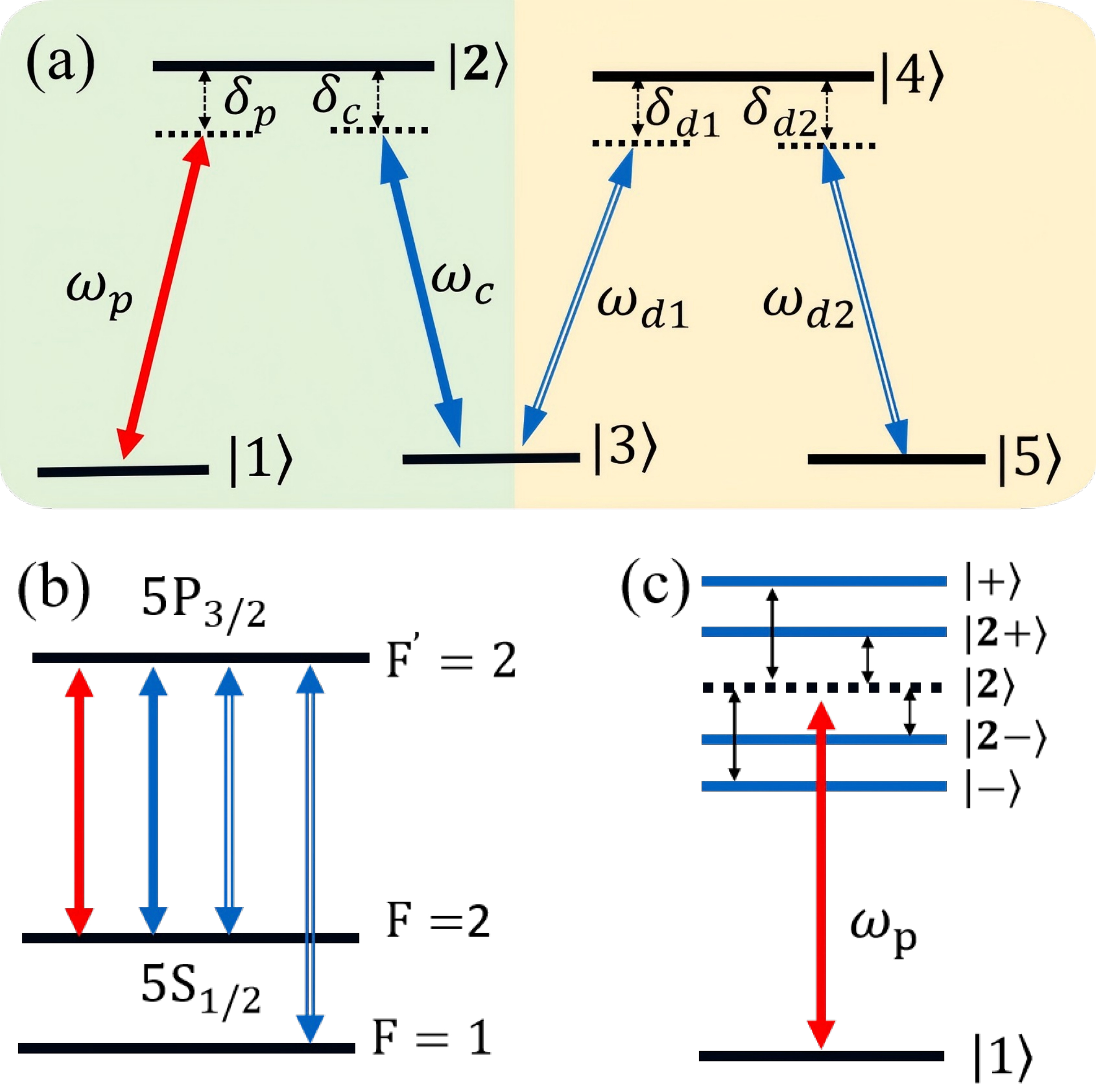}
\caption{\label{fig1}
(a) Schematic diagram of a five-level M-type atomic configuration coupled to a weak probe field and three strong fields.
(b) Relevant hyperfine levels of the $^{87}$Rb $D_2$ line corresponding to the proposed level scheme.
(c) Dressed-state picture, where the excited state $|2\rangle$ is coherently split into superposition branches ($|2\pm\rangle$, $|\pm\rangle$) gives rise to multiple interfering probe-coupling pathways.
The resulting interference reshapes the absorption spectrum and enables tunable spectral narrowing~\cite{Rubidium87,NewJPhys.26.043018}.}
\end{figure}

We consider an M-type configuration in a $^{87}$Rb atomic ensemble based on the $D_2$ line, as shown in Fig.~\ref{fig1}(a,b).
The specific Zeeman sublevels employed are
$|1\rangle=|5^2S_{1/2}$, $F=2, m_F=2\rangle$,
$|2\rangle=|5^2P_{3/2}$, $F'=2, m_F=1\rangle$,
$|3\rangle=|5^2S_{1/2}$, $F=2, m_F=0\rangle$,
$|4\rangle=|5^2P_{3/2}$, $F'=2, m_F=-1\rangle$,
and $|5\rangle=|5^2S_{1/2}$, $F=2, m_F=-2\rangle$.
A probe field with frequency $\omega_{p}$ and amplitude $E_{p}$ couples the transition $|1\rangle \leftrightarrow |2\rangle$,
while a control field with frequency $\omega_{c}$ and amplitude $E_{c}$ couples the transition $|2\rangle \leftrightarrow |3\rangle$.
Two assistant driving fields with frequency $\omega_{d1}$ and $\omega_{d2}$ (amplitude $E_{d1}$ and $E_{d2}$) couple the transitions $|4\rangle \leftrightarrow |3\rangle$ and $|5\rangle \leftrightarrow |4\rangle$, respectively, thereby introducing additional ground-state-mediated interference pathways beyond the conventional $\Lambda$ configuration.
The corresponding Rabi frequencies are defined as $\Omega_n=\mu_{ij}E_n/2\hbar$, where $\mu_{ij}$ is the relevant dipole moment for each transition.
Relevant spontaneous decays include $\Gamma_{45}=\Gamma_{43}$ (from $|4\rangle$) and $\Gamma_{21}=\Gamma_{23}$ (from $|2\rangle$).
As illustrated by the dressed-state picture in Fig.~\ref{fig1}(c), the strong fields hybridize the bare state $|2\rangle$ into multiple superposition branches, enabling multiple interfering pathways.
This interference modifies the dressed-state splittings and effective transition strengths, giving rise to tunable transparency windows with direction-dependent transmission, which form the basis for narrowband nonreciprocal propagation.

To quantitatively describe the interference dynamics discussed above, we formulate the interaction Hamiltonian for the five-level M-type system under the electric-dipole and rotating-wave approximations.
The resulting Hamiltonian reads
\begin{align}
\label{eq:H_Msystem}
\mathcal{H}_I/\hbar &= -\delta _p |2 \rangle \langle 2|  + ( \delta _{c} -\delta _p )|3 \rangle \langle 3| + ( \delta _{c} -\delta _p- \delta _{d1} ) |4\rangle \langle 4|
\nonumber\\
\nonumber\\
&+ ( \delta _{c} -\delta _p- \delta _{d1} + \delta _{d2} ) |5 \rangle \langle 5|
- [\Omega _p|1\rangle \langle 2|+ \Omega _{c}|2\rangle \langle 3|
\nonumber\\
\nonumber\\
& + \Omega _{d1}|3\rangle \langle 4| + \Omega _{d2}|4\rangle \langle 5|]
+ \text{H.c.},
\end{align}
where $\delta_{n} = \omega_{n}^{L}-\omega_{ij}$ is the single-photon detuning of the field from the the $\left| i \right\rangle \leftrightarrow \left| j \right\rangle$ transition.

Starting from the interaction Hamiltonian, we derive the dynamic equations (see Appendix~\ref{app.A}) for 25 density-matrix elements $\rho_{ij}$ ($i, j\in \{1, 2, 3, 4, 5\}$) from the master equation
\begin{equation}
\partial_t{\rho_{ij}}=-\frac{i}{\hbar}[\mathcal{H}_I,\rho_{ij}]+\mathcal{L}(\rho_{ij}),
\label{eq:master}
\end{equation}
where the Lindblad superoperator $\mathcal{L}(\rho_{ij})$ accounts for the population decay rates $\Gamma_{ij}$ and the coherence dephasing rates $\gamma_{ij}$.
In the presence of a weak probe field, such that the population remains mainly in $|1\rangle$, we solve the density-matrix equations under the steady-state condition ($\partial_t \rho_{ij} =0$) and obtain the probe coherence on the $\left\vert 1\right\rangle$ $\leftrightarrow$ $\left\vert 2\right\rangle$ transition,
\begin{equation}
\rho_{21}
=\frac{i\Omega_p}
{g_{21}+\frac{|\Omega_c|^{2}\mathcal{D}_{c}}{\mathcal{D}_{\mathrm{aux}}}},
\label{eq:rho21}
\end{equation}
with $\mathcal{D}_{\mathrm{aux}} \equiv g_{31} g_{41} g_{51} + |\Omega_{d1}|^{2} g_{51} + |\Omega_{d2}|^{2} g_{31}$ and $\mathcal{D}_{c} \equiv g_{41} g_{51} + |\Omega_{d2}|^{2}$.
Here $g_{ij}$ are the complex dephasing rates, generally written as $g_{ij}=\gamma_{ij}-i \Delta_{ij}$, where $\Delta_{ij}$ denotes the relevant single- or multi-photon detuning.
Their explicit expressions are listed in Appendix~\ref{app.A}.
This compact form highlights the structure of the five-level system in Fig.~\ref{fig1}(a).
The factor $\mathcal{D}_{\mathrm{aux}}$ collects the effective contribution of the assistant drivings $\Omega_{d1}$ and $\Omega_{d2}$ through states $\left\vert 3\right\rangle$, $\left\vert 4\right\rangle$, and $\left\vert 5\right\rangle$,
while $\mathcal{D}_{c}$ describes the response of the $\left\vert 4\right\rangle$ $\leftrightarrow$ $\left\vert 5\right\rangle$ branch modified by $\Omega_{d2}$.
Although $\mathcal{D}_{c}$ does not explicitly depend on the coupling field, it contributes to the probe coherence only through the product $|\Omega_c|^{2} \mathcal{D}_{c}$,which characterizes the effective feedback of the coupling field into this branch.
In the absence of the driving fields, Eq.~(\ref{eq:rho21}) reduces to the standard result for a three-level $\Lambda$ system under EIT conditions~\cite{RMP-EIT,jhwu-2009,PRAppl14.024032}.

We next examine the impact of Doppler broadening on the M-type atomic system.
To incorporate thermal motion, each single-photon detuning in Eq.~(\ref{eq:rho21}) is modified to include the velocity-dependent Doppler shift
\begin{equation}
\delta_{n} (v) =\delta_{n} - s_n\,k\,v,
\end{equation}
where $k$ is the common optical wavevector and $s_n=\pm1$ specifies whether the field co- or counter-propagates with z-axis.
Substituting these velocity-dependent detunings into the master equations gives the probe coherence $\rho_{21}(v)$ for each velocity class.
The probe susceptibility is obtained by averaging over the thermal velocity distribution,
\begin{equation}
\chi_p=\frac{N\mu_{21}^2}{\hbar\epsilon_0}\int f(v)\frac{\rho_{21}(v)}{\Omega_p}dv,
\label{eq:chi}
\end{equation}
where $f(v)=e^{- v^2/{u_p^2}}/u_p\sqrt{\pi}$ is the Maxwell-Boltzmann distribution.
Here, $N$ and $\epsilon_0$ denote the atomic density and vacuum permittivity, respectively, and $u_p=\sqrt{2k_{B}T/m}$ is the most probable atomic speed.
The probe absorption is proportional to Im $\chi_p$ which directly determines the transmission spectrum after thermal averaging.

While thermal motion modifies the single-photon detunings, an important qualification is that an approximate dark state survives only in the copropagating geometry.
The destructive-interference condition is therefore velocity-insensitive, and the M-type system supports a non-coupled superposition among the three ground states that underlies the transparency
\begin{equation}
\label{eq:dark}
|D_M\rangle = \frac{ {\Omega _{c}}{\Omega _{d2}}|1\rangle  - \Omega _{p}\Omega _{d2}|3\rangle  + \Omega _{p}\Omega _{d1}|5\rangle }{\sqrt {\Omega _{p}^2\Omega _{d1}^2 + \Omega _{p}^2\Omega _{d2}^2 + \Omega _{c}^2\Omega _{d2}^2}},
\end{equation}
where the coefficients are determined by the ratio of Rabi frequencies, such that the net dipole coupling to the excited states vanishes~\cite{CPB.2012,PhysRevA.109.043711}.
Analytical inspection shows that, in the copropagating configuration, the transparency dip remains close to two-photon resonance across all velocity classes, since Doppler shifts mainly influence the individual transitions but only weakly perturb their relative detuning.
This ensures that the interference condition sustaining $|D_M\rangle$ is preserved even after thermal averaging.
In the configuration with opposite propagation for the probe field, the cancellation becomes velocity dependent, and the dark-state interference is progressively washed out under Doppler averaging.

This theoretical picture connects the microscopic interference mechanism to the observed EIT feature in the macroscopic response, providing the foundation for the numerical analysis in the next section, where the influence of assistant fields on the EIT linewidth will be explored.

\section{Linewidth compression in Doppler-broadened atomic media}
\begin{figure}[ptbh]
\includegraphics[width=0.49\textwidth]{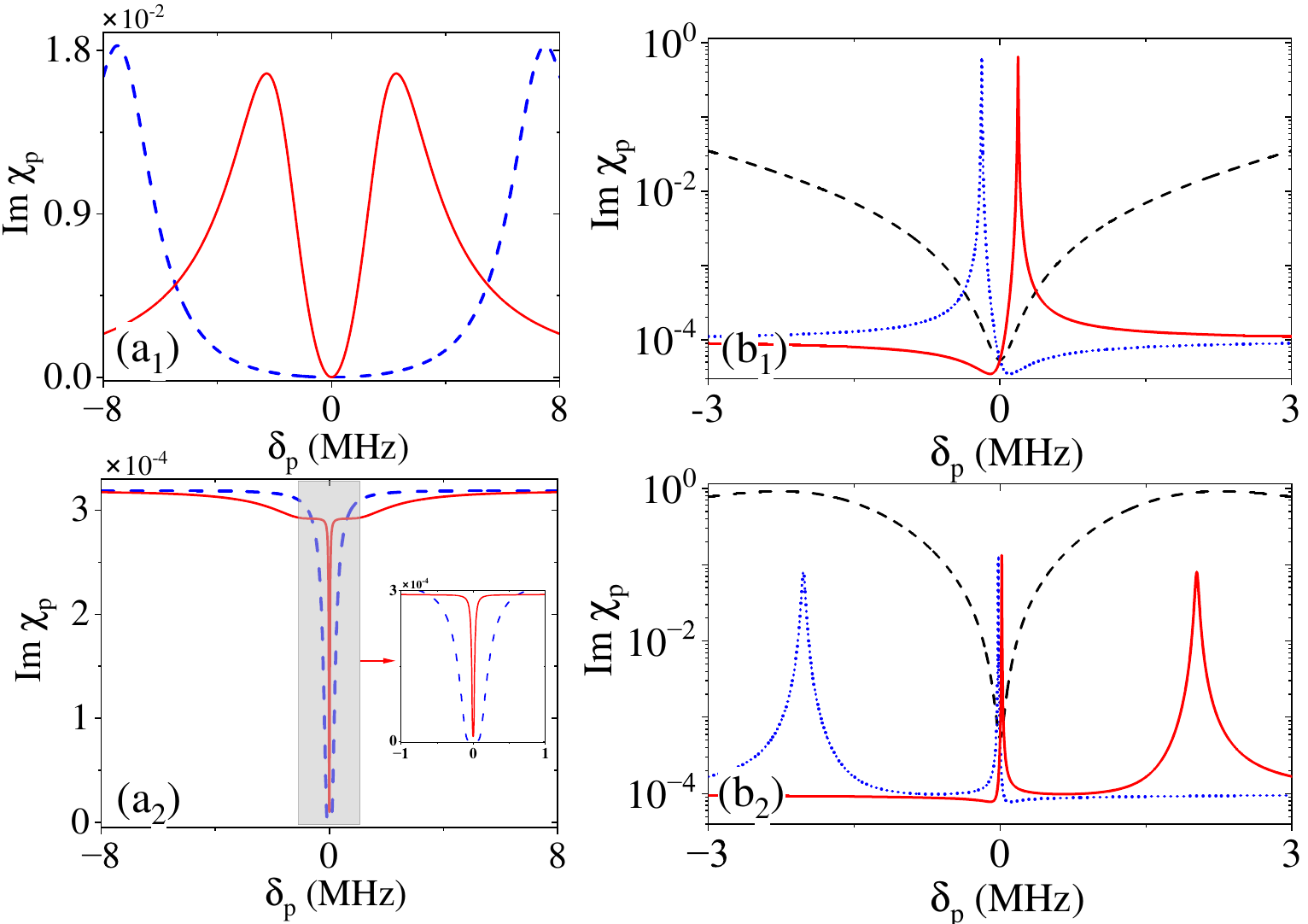}
\caption{\label{fig2}
 Imaginary part of the probe susceptibility $\mathrm{Im}[\chi_p]$ versus probe detuning $\delta_p$.
 Absorption profiles of the $\Lambda$-type (blue dashed) and M-type (red solid) systems at $T=0$ (a$_1$) and $300$\,K (a$_2$), respectively.
 Absorption profiles of the $\Lambda$-type (b$_1$) and M-type (b$_2$) systems, respectively, for selected atomic velocities: $v=0$ (black dashed), $+u_p$ (blue dotted), and $-u_p$ (red solid), where $u_p=\sqrt{2k_{\mathrm B}T/m}$ at $T=300$\,K.
 The shaded regions are enlarged in the insets.
The relevant parameters are: $\Omega _{d1}=3\Omega _{d2} =3\Omega _{c}= 2\pi \times 45$\,MHz, $\delta _{c} = \delta _{d1} = \delta _{d2} = 0$, $\Gamma _{21} =\Gamma _{23}= \Gamma _{43} =\Gamma _{45} = 2\pi \times 3$\,MHz, $\gamma_{12}=\gamma_{14}=\gamma_{23}=\gamma_{25}=\gamma_{34}=\gamma_{45}=\gamma_{24}/2=2 \pi \times 3$\,MHz, $\gamma_{13}=\gamma_{15}=2 \pi \times 1$\,kHz, and $N=1\times10^{12}$\,cm$^{-3}$.}
\end{figure}

We first compare the probe absorption profiles of the M-system and a standard $\Lambda$-system.
At $T=0$\,K [Fig.~\ref{fig2}(a$_1$)], the central transparency window ($\delta_p=0$) in the M-system has a FWHM about five times smaller than that in the $\Lambda$-system.
At $T=300$\,K [Fig.~\ref{fig2}(a$_2$)], Doppler averaging increases this disparity, and the M-system window is nearly two orders of magnitude narrower.
To examine how different atomic velocities contribute to this behavior, Figs.~\ref{fig2}(b$_1$) and \ref{fig2}(b$_2$) show the responses of the $\Lambda$-type and M-type systems for $v=0$ and $v=\pm u_p$, where $u_p$ is the most probable speed at 300\,K. The $v=0$ curves retain the distinct absorption structures seen at T=0\,K.
For $v=\pm u_p$, the spectral features shift in opposite directions about $\delta_p=0$.
These velocity-dependent peaks lie much closer to resonance in the M-type system [Fig.~\ref{fig2}(b$_2$)] than in the $\Lambda$-type system [Fig.~\ref{fig2}(b$_1$)].
%The insets show that the absorption features nearest the central transparency point lie closer to it in the M-type system than in the $\Lambda$-type system.
This comparison provides a qualitative explanation for the much narrower transparency window of the M-type system after Doppler averaging at $T=300$ K [Fig.~\ref{fig2}(a$_2$)].
We next analyze the dressed states to clarify how the two assistant driving fields in the M-type system modify the velocity-dependent resonances.

\subsection{Spectral narrowing from the dressed-state perspective}
The spectral features in Fig.~\ref{fig2} (a$_1$, a$_2$) can be further understood within a dressed-state framework~\cite{PRA.83.033830,CPB.2012,PRA.98.013423,NewJPhys.26.075002}.
The two assistant drivings $(\Omega_{d1},\Omega_{d2})$ open additional excitation channels that interfere with the conventional $\Lambda$-scheme and give rise to a modified dressed-state spectrum.
Diagonalization of the interaction Hamiltonian yields four dressed eigenenergies $\lambda_i(v)$ $(i=1,\dots,4)$, which can be classified into two inner states $(\lambda_{1,2})$ with dominant $|2\rangle$ component and two outer states $(\lambda_{3,4})$ largely formed by $|3\rangle$, $|4\rangle$, and $|5\rangle$.
For compactness, we introduce the generalized Rabi frequencies
\begin{equation}
\Lambda_\pm^2=2(\Omega_c^2+\Omega_{d1}^2+\Omega_{d2}^2)\,\pm\,2\sqrt{(\Omega_c^2+\Omega_{d1}^2+\Omega_{d2}^2)^2-4\Omega_c^2\Omega_{d2}^2},
\label{eq:Lambda-pm}
\end{equation}
so that the eigenenergies can be written as
\begin{equation}
\begin{aligned}
\lambda_{1,2}(v)&=\frac{kv\pm\sqrt{(kv)^2+\Lambda_-^2}}{2},
\\
\lambda_{3,4}(v)&=\frac{kv\pm\sqrt{(kv)^2+\Lambda_+^2}}{2}.
\label{eq:lambdas-compact}
\end{aligned}
\end{equation}

\begin{figure}[ptbh]
\includegraphics[width=0.48\textwidth]{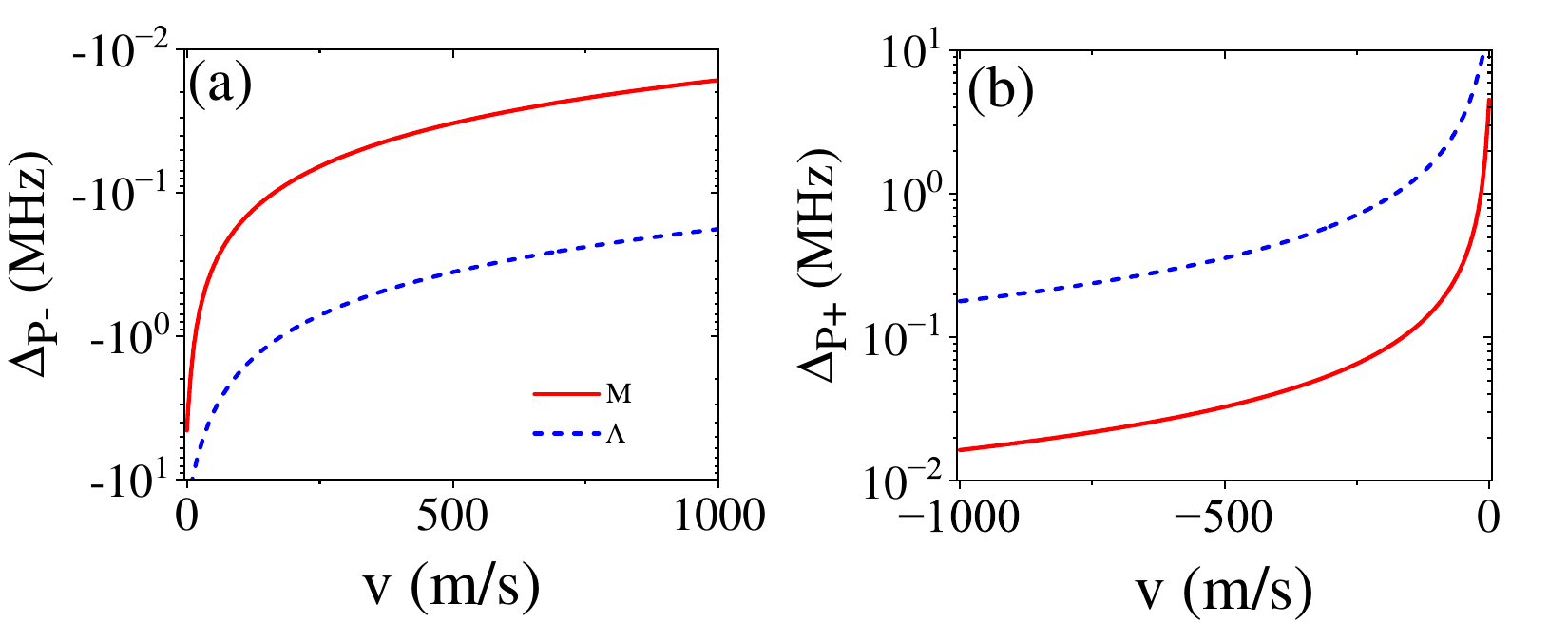}
\caption{\label{fig-Position}
(a) Left absorption peak $\Delta _{p-}$ versus positive velocity and (b) right peak $\Delta _{p+}$ versus negative velocity for the M-system (red solid line) and $\Lambda$-system (blue dashed line).
Parameters are as in Fig.~\ref{fig2}. }
\end{figure}

Physically, the inner pair $(\lambda_{1,2})$ is most relevant to the probe transparency window, as the corresponding dressed states retain a substantial contribution from $|2\rangle$ and give rise to the two absorption resonances closest to the line center.
In contrast, the outer pair $(\lambda_{3,4})$ originates from dressed couplings involving the auxiliary manifold formed by $|3\rangle$, $|4\rangle$, and $|5\rangle$, and mainly contributes to absorption features away from resonance.
For a velocity class $v$, the effective transparency linewidth can be associated with the splitting of the inner pair,
\begin{equation}
\Gamma_{\mathrm{EIT}}^{(M)}(v)=|\lambda_1(v)-\lambda_2(v)|=\sqrt{(kv)^2+\Lambda_-^2},
\label{eq:GammaEIT-M}
\end{equation}
where $\Lambda_-$ denotes the inner-state splitting at zero velocity.
Numerical simulations in Fig.~\ref{fig-Position} show that increasing \(|v|\) shifts the inner eigenenergies asymmetrically with respect to the probe resonance, thereby reducing \(\Gamma_{\mathrm{EIT}}^{(M)}\) and narrowing the transparency window.
Thermal averaging over velocity classes enhances this reduction, because different velocity classes overlap near resonance, which increases absorption and further compresses the transparency feature.
Consequently, the M-system exhibits a sharper transparency profile than the conventional $\Lambda$-system in Fig.~\ref{fig2}, as the two assistant drivings enhance the velocity sensitivity of the dressed-state splitting.
In realistic ensembles, the finite ground-state coherence time imposes a lower bound on the achievable linewidth~\cite{PhysRevA.51.576,PhysRevA.65.022314,PhysRevA.72.069905,PhysRevR.8.L032046}.
This practical limitation, however, does not affect the dressed-state narrowing mechanism discussed above.

\subsection{Comparison with the $\Lambda$ configuration}
To establish the connection with the standard three-level case, we examine the limit where both assistant driving fields vanish ($\Omega_{d1}$,$\Omega_{d2}\to0$). In this regime, the outer pair of eigenenergies ($\lambda_{3,4}$) smoothly reduces to the well-known $\Lambda$-type result,
\begin{equation}
\lambda_{\Lambda,\pm}(v)=\frac{kv\pm\sqrt{(kv)^2+4\Omega_c^2}}{2},
\label{eq:lambda-Lambda}
\end{equation}
recovering the usual transparency splitting $2\Omega_c$. The inner pair ($\lambda_{1,2}$) collapses to $\{0,kv\}$, corresponding to one resonant state and one purely Doppler-shifted state without forming a finite transparency window.

Direct comparison in Fig.~\ref{fig-Position} reveals that, for any given velocity class, the M-system resonances approach the probe resonance more closely than their $\Lambda$-system counterparts. The assistant drivings $\Omega_{d1}$ and $\Omega_{d2}$ introduce new interference channels that increase the velocity sensitivity of the dressed eigenenergies, strongly compressing the transparency window. Thermal velocity averaging further amplifies this effect, explaining the significantly narrower observed linewidth of the M-system.

When ($\Omega_{d1}\gg\Omega_{c,d2}$), the dressed levels are shifted far from the probe resonance, and the transparency window narrows $\Gamma_{\mathrm{EIT}}^{(M)}\to 0$.
In the opposite limit, $\Omega_{d1},\Omega_{d2}\to 0$, the system continuously approaches the standard $\Lambda$-type behavior.
However, this suppression relies on symmetric driving and ideal coherence conditions, and any imbalance $\Omega_{d1}\neq \Omega_{d2}$ or additional dephasing channels $\gamma_{ij}$ restores the usual EIT splitting.
Moreover, if the beams are not perfectly co-propagating, uncompensated two-photon Doppler shifts enter the Hamiltonian, breaking this narrowing mechanism.
These predictions are experimentally testable and define the parameter space where linewidth suppression can be observed.

If the probe propagates in the opposite direction, the Doppler-cancellation condition breaks down, and the transparency window collapses into strong broadband absorption.
This provides a natural pathway to engineer direction-dependent narrowband transmission, forming the basis of the nonreciprocal effects analyzed in Sec.~\ref{sec:nonreciprocity}.

\section{\label{sec:nonreciprocity} Directional transparency underlying narrowband optical nonreciprocity}

Building on the above results, we next examine how breaking the co-propagating configuration affects the probe transmission and how assistant driving fields can modulate this nonreciprocity~\cite{NewJPhys.26.043018}.
When the probe field propagates in the opposite direction to the other strong fields, the Doppler compensation that previously enabled a narrow transparency window is disrupted.
As a result, atoms in different velocity classes experience significant detuning, leading to strong absorption even at the same spectral position where full transparency was observed in the co-propagating case~\cite{PRAppl18.024027,SciAdv7.abe8924,CommunPhys.6.33,Nat.Phot12,LPR17.2022}.

The directional dependence of the probe transmission $T_p^{\pm}$ manifests as a nonreciprocal response in the M-type system, where the forward ($+$) and backward ($-$) transmissions are given by
\begin{equation}
T_p^{\pm} = \exp[-k_p L\, \mathrm{Im}\chi_p^{\pm}].
\end{equation}
Here $\chi_p^{\pm}$ denotes the probe susceptibility in the two propagation directions, $k_p$ is the probe wave vector, and $L$ is the length of the medium.

\begin{figure}[ptbh]
\includegraphics[width=0.48\textwidth]{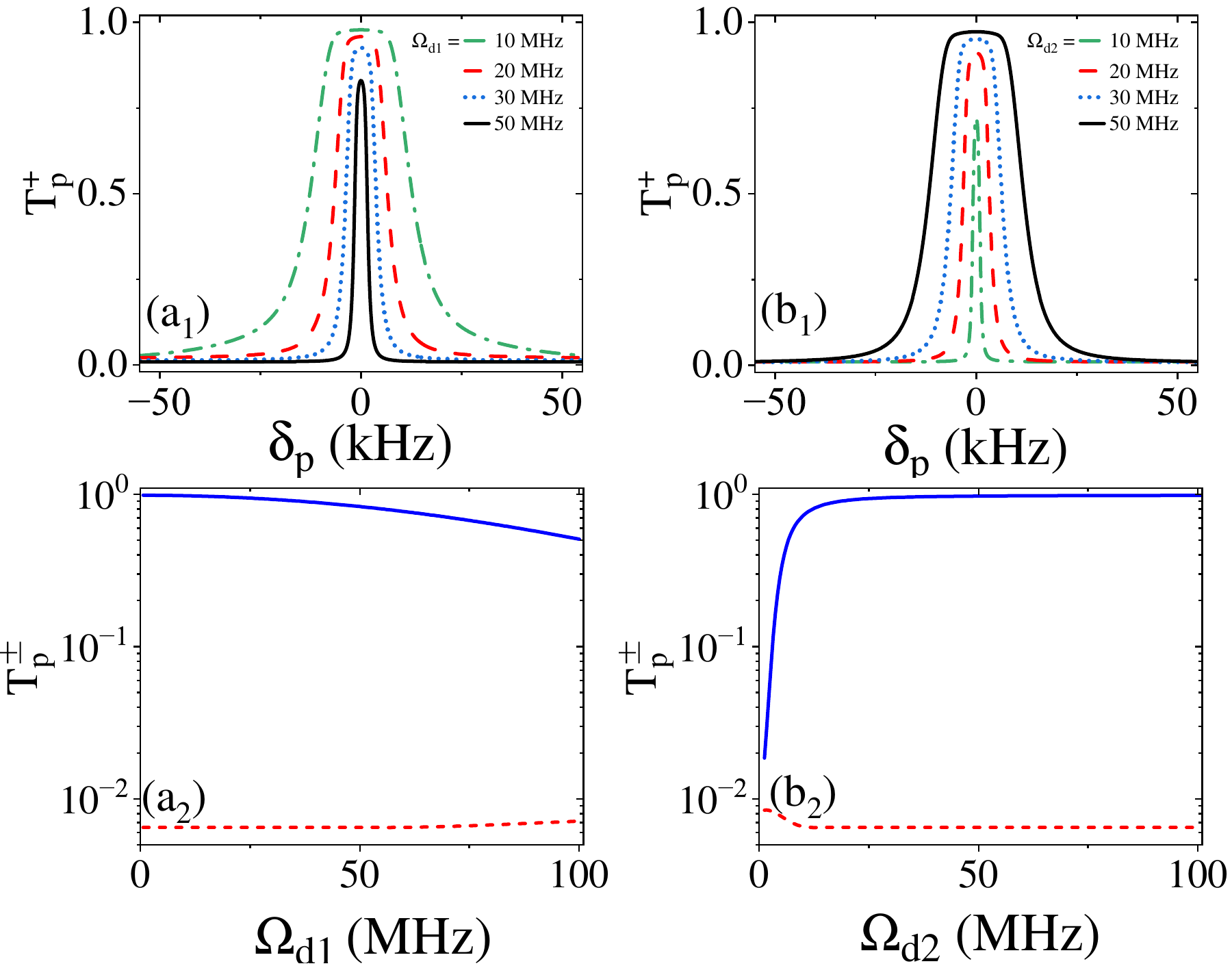}
\caption{\label{Fig-transmission}
Forward probe transmission spectra $T_p^+$  as a function of probe detuning $\delta_{p}$  for different values of $\Omega_{d1}$ (a$_1$) and $\Omega_{d2}$ (b$_1$), respectively.
Probe transmission at resonance ($\delta_p=0$) for forward and backward propagation as a function of $\Omega_{d1}$ (a$_2$) and $\Omega_{d2}$ (b$_2$).
Relevant parameters are the same as in Fig.~\ref{fig2} except $L=2$\,mm.
}
\end{figure}

Furthermore, the presence of the assistant driving fields $\omega_{d1}$ and $\omega_{d2}$ offers a means to actively tune this nonreciprocity.
As shown in Fig.~\ref{Fig-transmission}(a$_1$,a$_2$), increasing $\Omega_{d1}$ narrows the absorption profile and simultaneously reducing the on-resonance forward transmission, though it remains above $82\%$,
while the backward transmission stays close to zero (not shown), ensuring strong directionality.
In contrast, increasing $\Omega_{d2}$ broadens the transparency window and further enhances the forward transmission beyond $90\%$, still maintaining negligible backward transmission, as illustrated in Fig.~\ref{Fig-transmission}(b$_1$,b$_2$).

\begin{figure}[ptbh]
\includegraphics[width=0.48\textwidth]{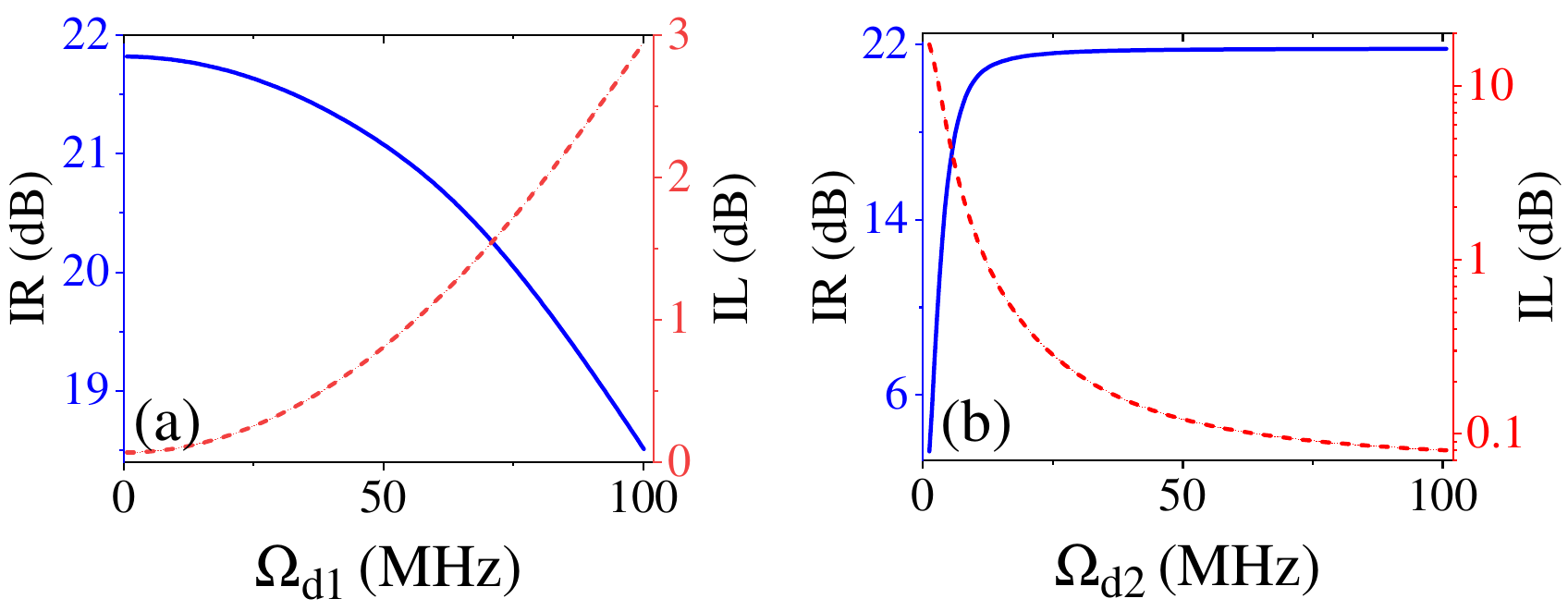}
\caption{\label{Fig-RL}
Isolation ratio IR (a) and insertion loss IL (b) against $\Omega_{d1}$ (a) and $\Omega_{d2}$ (b).
Relevant parameters are the same as in Fig.~\ref{Fig-transmission}. }
\end{figure}
Fig.~\ref{Fig-RL} quantifies the isolation performance under these conditions.
The isolation ratio (IR) and insertion loss (IL) are defined as
\begin{equation}
\mathrm{IR}=10\log_{10}\Big(\frac{T_p^{+}}{T_p^{-}}\Big), \qquad
\mathrm{IL}=-10\log_{10}T_p^{+},
\end{equation}
where $T_p^{\pm}$ denote the forward and backward probe transmissions, respectively.
Panel (a) shows that increasing $\Omega_{d1}$ slightly reduces the isolation ratio (IR, blue solid line, left axis) while insertion loss (IL, red dashed line, right axis) gradually increases, exceeding 1\,dB for $\Omega_{d1}>57$\,MHz.
This behavior arises because stronger coupling of the inner dressed states increases residual absorption in the forward direction, degrading isolation efficiency.
In contrast, increasing $\Omega_{d2}$ enhances IR to nearly 22\,dB and reduces IL to below 0.1\,dB, indicating improved nonreciprocal transmission.
The spectral linewidth also broadens, as shown in Fig.~\ref{Fig-transmission}(b$_1$), thereby reducing frequency selectivity.
Physically, the two assistant driving fields influence the interference among dressed states in distinct ways, leading to different modifications of the nonreciprocal transmission.
Increasing $\Omega_{d1}$ enhances residual absorption in the forward direction, slightly degrading the isolation ratio and increasing insertion loss.
In contrast, increasing $\Omega_{d2}$ suppresses forward loss and improves isolation, albeit with a broader spectral response, highlighting two different routes for engineering the trade-off between nonreciprocal efficiency and spectral selectivity.

\section{conclusion}
In conclusion, we have theoretically investigated narrowband nonreciprocal transmission and its tunability in a hot $^{87}$Rb vapor with an M-type level configuration. %atomic M-system.
Coherent interference gives rise to a narrow EIT resonance, while a Doppler-insensitive two-photon resonance allows this resonance to persist after velocity averaging for forward probe propagation.
For backward probe propagation, the two-photon resonance becomes Doppler sensitive. This leads to strong absorption and nearly vanishing transmission.
By contrast, the forward transmission remains above 80\%.
Together, these results establish pronounced nonreciprocal transmission in the thermal vapor.
The two assistant driving fields couple to distinct excitation pathways and provide further control over the nonreciprocal response.
Specifically, increasing $\Omega_{d1}$ narrows the transparency linewidth, but also increases the insertion loss beyond 1\,dB at large $\Omega_{d1}$ and reduces the isolation ratio.
In contrast, increasing $\Omega_{d2}$ raises the isolation ratio to approximately 22\,dB while keeping the insertion loss below 0.1\,dB, although the nonreciprocal bandwidth becomes broader.
The opposite trends produced by the two fields allow the spectral width and nonreciprocal transmission performance to be adjusted within the same system.
Overall, our results show that propagation geometry and multilevel interference enable a narrow transmission window with high forward transmission and strong reverse suppression in a Doppler-broadened thermal vapor, offering a route to combining spectral selectivity with optical isolation.

\section*{ACKNOWLEDGMENTS}
This work is supported by the National Natural Science Foundation of China (Grants Nos.~12604536, 12404405, and 11575051),
the Hebei Natural Science Foundation (Grants Nos.~A2025205028 and A2021205020),
and the Science Foundation of Hebei Normal University of China (Grants Nos.~L2025B11 and L2026J02). Y.K.B. also acknowledges support from Hebei 333 Talent Project (No. B20231005).

\appendix
\begin{widetext}
\section{\label{app.A} Density-matrix equations for M system}
Here, we derive the density-matrix equations for the five-level M-type system shown in Fig.~\ref{fig1}. Substituting the interaction Hamiltonian (\ref{eq:H_Msystem}) into the master equation (\ref{eq:master}), we obtain
\begin{equation}\label{eqs-0}
\begin{aligned}
\partial _t\rho _{11} &= \Gamma _{21}\rho _{22} - i\Omega _{p}\rho _{12} + i\Omega_{p}^*\rho _{21},\\
\partial _t\rho _{22} &= -\Gamma _{21}\rho _{22}-\Gamma _{23}\rho _{22} - i\Omega_{p}{\rho_{21}} + i\Omega _{p}^*\rho_{12}- i\Omega_{c}^*\rho _{23}
+ i\Omega _{c}\rho _{32},\\
\partial _t\rho _{33} &= \Gamma _{23}\rho _{22}+\Gamma _{43}\rho _{44} - i\Omega_{c}\rho_{32} + i\Omega _{c}^*\rho_{23}- i\Omega_{d1}^*\rho _{43}
+ i\Omega _{d1}\rho _{34},\\
\partial _t\rho _{44} &=  - \Gamma _{43}\rho _{44}- \Gamma _{45}\rho _{44} - i\Omega_{d1}\rho _{43} + i\Omega _{d1}^*\rho _{34} + i\Omega _{d2}\rho _{54}
 - i\Omega _{d2}^*\rho _{45},\\
\partial _t\rho _{12} &= -g_{12}\rho _{12} - i\Omega _{p}\left( \rho _{11} - \rho _{22} \right)- i\Omega _{c}^*\rho _{13},\\
\partial _t\rho _{13} &= -g_{13}\rho _{13} - i\Omega _{c}\rho _{12}- i\Omega _{d1}^*\rho _{14} + i\Omega _{p}\rho _{23},\\
\partial _t\rho _{14} &=  -g_{14}\rho _{14}- i\Omega _{d1}\rho _{13} - i\Omega _{d2}^*\rho _{15} + i\Omega _{p}\rho _{24},\\
\partial _t\rho _{15} &= - g_{15}\rho _{15}- i\Omega _{d2}\rho _{14} + i\Omega _{p}\rho _{25},\\
\partial _t\rho _{23} &= -g_{23}\rho _{23} - i\Omega _{c}\left( \rho _{22} - \rho _{33} \right)- i\Omega _{d1}^*\rho _{24} + i\Omega _{p}^*\rho _{13},\\
\partial _t\rho _{24} &= -g_{24}\rho _{24} + i\Omega _{c}\rho _{34}- i\Omega _{d1}\rho _{23} + i\Omega _{p}^*\rho _{14}
- i\Omega _{d2}^*{\rho _{25}},\\
\partial _t\rho _{25} &= -g_{25}\rho _{25} + i\Omega _{c}\rho _{35}+ i\Omega _{p}^*\rho _{15} - i\Omega _{d2}\rho _{24},\\
\partial _t\rho _{34} &= -g_{34}\rho _{34} - i\Omega _{d1}\left(\rho _{33} - \rho _{44} \right)+ i\Omega _{c}^*\rho _{24} - i\Omega _{d2}^*\rho _{35},\\
\partial _t\rho _{35} &= -g_{35}\rho _{35} + i\Omega _{d1}\rho _{45}+ i\Omega _{c}^*\rho _{25} - i\Omega _{d2}\rho _{34},\\
\partial _t\rho _{45} &= -g_{45}\rho _{45} - i\Omega _{d2}\left( \rho _{44} - \rho _{55}\right)+ i\Omega _{d1}^*\rho _{35}.
\end{aligned}
\end{equation}
These equations are supplemented by the normalization condition $\sum_i\rho _{ii}=1$ and $\rho_{ij}=\rho_{ji}^{\ast}$.
The complex coefficients are defined as $g _{21} = \gamma_{21}-i \delta_{p}$, $g _{31}= \gamma _{31}-i (\delta_{p} - \delta_{c})$, $g _{41} = \gamma_{41}-i (\delta_{p} - \delta_{c} + \delta_{d1})$,
and $g _{51} = \gamma_{51}-i (\delta_{p} - \delta_{c} + \delta_{d1} - \delta_{d2})$, with $g_{ij}=g_{ji}^{\ast}$.
Here, $\gamma _{31}(\gamma _{51})$ denote the collisional decay rate associated with the $\left\vert 1\right\rangle$ $\leftrightarrow$ $\left\vert 3\right\rangle$ ($\left\vert 1\right\rangle$ $\leftrightarrow$ $\left\vert 5\right\rangle$) transition.

Under the approximations specified in the main text, the steady-state solution of Eq.~(\ref{eqs-0}) yields the probe coherence $\rho_{21}$ given in Eq.~(\ref{eq:rho21}).

\vspace{2mm}

\end{widetext}

\end{document}